# Pick-up and assembling of chemically sensitive van der Waals heterostructures using dry cryogenic exfoliation

Vilas Patil[1], Sanat Ghosh[1], Amit Basu[1], Kuldeep[1], Achintya Dutta[1], Khushabu Agrawal[1], Neha Bhatia[1], Amit Shah[1], Digambar A. Jangade[1], Ruta Kulkarni[1], A. Thamizhavel[1] and Mandar M. Deshmukh[1*]

[1]Department of Condensed Matter Physics and Materials Science, Tata Institute of Fundamental Research, Colaba, Mumbai 400005

**Abstract**

Assembling atomic layers of van der Waals materials (vdW) combines the physics of two materials, offering opportunities for novel functional devices. Realization of this has been possible because of advancements in nanofabrication processes which often involve chemical processing of the materials under study; this can be detrimental to device performance. To address this issue, we have developed a modified micro-manipulator setup for cryogenic exfoliation, pick up, and transfer of vdW materials to assemble heterostructures. We use the glass transition of a polymer PDMS to cleave a flake into two, followed by its pick-up and drop to form pristine twisted junctions. To demonstrate the potential of the technique, we fabricated twisted heterostructure of $Bi_2Sr_2CaCu_2O_{8+x}$ (BSCCO), a van der Waals high-temperature cuprate superconductor. We also employed this method to re-exfoliate $NbSe_2$ and make twisted heterostructure. Transport measurements of the fabricated devices indicate the high quality of the artificial twisted interface. In addition, we extend this cryogenic exfoliation method for other vdW materials, offering an effective way of assembling heterostructures and twisted junctions with pristine interfaces.

*deshmukh@tifr.res.in

**Introduction :**

Two-dimensional (2D) vdW materials have great potential, providing opportunities for scientific and technological innovation[1-3]. Various van der Waals -heterostructures (vdWH) have been fabricated by stacking different 2D materials like graphene, hexagonal boron nitride (h-BN), and other 2D transition metal dichalcogenides (TMDCs) for the study of fundamental electronic[4,5] and optical properties[6,7]. vdWH provide a unique avenue to bring materials with completely different, sometimes even with competing properties, together in a single artificial material system. The device fabrication with 2D materials mainly includes two approaches. In bottom-up approach, growth of thin films of different materials are carried out using techniques like chemical vapour deposition[8] and epitaxial growth[9]. The top-down approach, on the other hand, consists of growing bulk single crystals followed by mechanical exfoliation[10] or anodic bonding exfoliation[11,12], among other methods to thin it down to atomic thicknesses.

Fabrication of heterostructures using mechanically exfoliated flake uses various transfer methods such as wet transfer[13], dry released transfer[14] and polycarbonate[15] (PC)/polypropylene carbonate (PPC)[16] based stamping process. Most processes use the dry transfer and PC/PPC-based stamping methods as these techniques are flexible and efficient for fabricating various types of heterostructures. Usually, the dry transfer method uses a polymer polydimethylsiloxane (PDMS)[14]. In this method, 2D materials are exfoliated on this polymer sheet and then transferred onto other substrates. Although this dry transfer method does not use any organic solvent/chemical it is known to leave residues on the transferred material surface. The stamping method uses PPC and PC polymeric thin films to pick up and drop 2D materials and make heterostructure of different 2D materials. An organic solvent is required at the final step of device fabrication for dissolving PPC and PC polymer thin films[15-17]. Although the heterostructures are typically encapsulated between hBN and do not come in direct contact with the organic solvent, there are possibilities of the solvents seeping through the heterostructure; this can be detrimental for various chemically sensitive materials. Among many, one such material is the high-temperature cuprate superconductor BSCCO. As an example, we now describe the fabrication of twisted devices of BSCCO using the cryogenic dry exfoliation technique. We also study the efficacy of this technique for other 2D materials, including $CrCl_3$, $NbSe_2$, $WSe_2$, $MoS_2$ and $GeBi_2Te_4$.

There is an increasing interest in studying the unusual superconducting properties of a few atom-thick van der Waals superconductors. Among the layered superconductors, BSCCO has garnered significant attention recently[11,12,18]. Studying the superconducting properties of 2D BSCCO yielded inconsistent electrical response[19] due to the degradation in properties of few-atom thin BSCCO in an ambient atmosphere. The degradation of superconductivity results in out-diffusion of oxygen atoms from the lattice above 200 K[20,21]. However, covering the material with graphene/h-BN has been shown to slow the process of degradation[18,20,22]. Also, the chemically unstable nature of BSCCO quickly leads to surface deterioration by the adsorption of water molecules[22-24]. As a result, the fabrication of heterostructures involving BSCCO using the standard PPC/PC and PDMS stamp methods has limitations in controlling these conditions, and the fabrication of twisted BSCCO devices has proven to be challenging[20]. Here we employ cryogenic dry exfoliation in a controlled environment to overcome these difficulties. This method ensures high-quality superconducting devices by avoiding any chemical processing. Our technique provides a unique opportunity to address and solve fabrication-related problems with BSCCO. The cryogenic pick-up and transfer method we have developed is motivated by the work of Zhao et al.[25] and improves upon it. We realize more control of the meniscus by using the hemispherical PDMS stamps[26] and exploit the glass transition of PDMS at -120 °C to exfoliate the crystal rather than using external mechanical force. Avoiding external force for cleavage reduces the chances of cracking and microscopic damage to the flakes. Devices have been fabricated by re-exfoliating a bulk BSCCO flake into two pieces and stacking them using a PDMS stamp; this ensures alignment of the crystal axis and a well-defined twist angle (with an accuracy of ± 0.5°). The time for device fabrication has also been optimized for better junction properties. We demonstrate that the devices fabricated by this method exhibit high quality, as suggested by the superconducting transition temperature and critical current density of the junction. Further, the developed method can be utilized for other 2D chemically sensitive materials.

**Cryogenic setup details:**

The low-temperature exfoliation was carried out inside a glove box in an argon (Ar) atmosphere with $O_2$ and $H_2O$ levels < 0.1 ppm. We accomplished better repeatability of devices using Argon compared to our initial attempts using nitrogen in the glove box. The low-temperature stage inside the glove box used for cryogenic exfoliation is custom-made and is depicted in Figure 1a. An existing 2D flake transfer setup has been modified for cryogenic exfoliation, as presented in Figure 1b-c. The cold stage, utilized for exfoliation at low temperatures, is a stainless-steel hollow cylinder with a top copper surface measuring 2.5 cm in diameter and 2.5 cm in height. Two Polytetrafluoroethylene (PTFE) tubes are connected to the stage to circulate liquid nitrogen ($N_2$), which cools the substrate. One end of the PTFE tube is connected to the liquid $N_2$ container, while the other is connected to a vacuum pump via a valve to control the flow of the cryogen to the stage. Controlling the impedance of the valve controls the flow of cryogen. The temperature sensor mounted on the cold stage monitors the sample temperature. Additionally, the cold stage is attached to a micromanipulator to enable sample movement in X, Y, and Z directions. During cryogenic exfoliation, we keep the substrate with mechanically exfoliated BSCCO flake on the stage and start a vacuum pump for circulating liquid $N_2$ through the PTFE tube for cooling down the stage. More details of the low-temperature exfoliation procedure are described next.

**Device Fabrication Methods:**

The cryogenic exfoliation technique enables the re-exfoliation of a thicker BSCCO flake, which was previously mechanically exfoliated on a $SiO_2$/Si chip using scotch tape. The re-exfoliated flake is then immediately transferred over the remaining flake on the substrate to form the twisted junction, as shown in Figure 2, while maintaining the cryogenic temperature in the controlled environment. This low-temperature exfoliation method effectively prevents the degradation of the interfacial and superconducting properties of BSCCO crystals.

We use a hemispherical PDMS stamp[26] as this shape accurately controls meniscus flow for preferential flake pick-up and transfer during 2D heterostructure fabrication. First, we cut the commercially available Gel-pak PDMS (PF-60x60-0065-X4) into cylindrical pieces with a diameter of ~2 mm. We then put a small drop of homemade viscous PDMS onto the cylindrical pieces to achieve the hemispherical shape and cure it at 150 °C for 15 min. The homemade PDMS is prepared using a silicone elastomer base and curing agent Sylgard 184 (Sigma Aldrich) with a 10:1 ratio. For the exfoliation of BSCCO crystal, we used a $SiO_2$/Si substrate, which was cleaned by $O_2$ plasma for 90 sec and heated at 150 °C overnight inside the glove box. First, we mechanically exfoliate BSCCO (Bi-2212) crystal with scotch tape and transfer the exfoliated BSCCO flake onto the pre-cleaned $SiO_2$/Si substrate. Then we put the substrate on the cryogenic stage and search for thick (~50-80 nm) and uniform BSCCO flakes through the optical microscope. Once we identify the desired flake, we cool down the cryogenic setup to -25 °C using liquid nitrogen. We then attach the hemispherical PDMS stamp to the BSCCO flake and wait for the temperature to reach ~ -120 °C. At this temperature, the PDMS goes through a glass transition. Because of the transition, the PDMS meniscus hardens and cleaves off a part of the BSCCO flake. The cleaved flake gets attached to the PDMS stamp. Then we quickly rotate the PDMS stamp at an angle of choice and re-attach the PDMS stamp on top of the bottom flake, as shown in Figure 2. We aim to minimize the time within a minute between the re-exfoliation and stacking process for BSCCO. We stop the cooling and wait for the stage temperature to rise above 10° C and slowly remove the hemispherical PDMS stamp. At this temperature, the adhesion of the BSCCO flake with PDMS becomes low, resulting in the detachment of the flake from PDMS. Finally, we align a SiN mask with the fabricated stack and deposit gold (Au) metal electrodes using e-beam evaporation. Thus, our entire fabrication process does not involve any chemical processing of the device and does not require mechanical strain. After completing device fabrication, we immediately load the device in a high vacuum ($2.5\times10^{-6}$ mbar) cryostat for measurement.

**Electrical characterization of fabricated twisted BSCCO devices:**

To check the quality of fabricated devices, we perform electrical transport measurements using a four-probe configuration in a Lakeshore cryogenic probe station (Lakeshore CRX-6.5K) for a nominally zero-degree twisted BSCCO device shown in Figure 3a and 4a. The four-probe resistance as a function of temperature is shown in Figure 3b. The blue and pink curves are the response from the twisted junction and pristine BSCCO, respectively. Both curves exhibit superconducting transition temperatures close to each other. The measured critical temperature $T_c$ is 86 K, consistent with the reported values for the Bi-2212 phase of the BSCCO crystal[21]. Additionally, a single $T_c$ of junction suggests that the low-temperature re-exfoliated BSCCO flake contains a single phase, and the interface is clean. We tested several devices[27] and found that their $T_c$ was similar. Figure 3c shows the current-voltage characteristics (IVCs) across the junction at 70 K. There are multiple discrete jumps in the IV characteristics in the superconducting-to-normal state transition. The first jump corresponds to the critical current of the interfacial Josephson junction whereas the consecutive jumps at higher currents come from switching of intrinsic Josephson junctions along the c-axis which are inherent to the bulk BSCCO crystal[27]. The critical current $I_c$ of the interfacial junction at 70 K is 1.2 mA, which translates to a critical current density $J_c$ of 0.34 kA/cm$^2$, as shown in Figure 3c. Figure 3d shows the dc I-V characteristics for the temperature range of 70 K to 90 K. In Figure 4b we observe the critical current density of a nominally 0-degree twisted junction at 10 K is ~ 1.34 kA/cm$^2$ which is comparable to the critical current density of the intrinsic Josephson junctions in the bulk BSCCO[28]; this is direct evidence of the high-quality nature of the junctions made by the cryogenic exfoliation method.

**Pick-up and drop of other vdW materials:**

We tested the versatility of our modified cryogenic exfoliation methodology by fabricating heterostructures with a wide range of 2D materials, including $CrCl_3$, $NbSe_2$, $MoS_2$, $WSe_2$, and $GeBi_2Te_4$. These materials were initially exfoliated onto a pre-cleaned $SiO_2$/Si substrate, followed by the re-exfoliation step described previously. As shown in Figure 5a and b, $CrCl_3$ and $NbSe_2$ flakes with a thickness of approximately 60 and 130 nm were successfully re-exfoliated and dropped on the original flake to fabricate a nominally 0º twisted structure. Implementing our modified low-temperature re-exfoliation method in an Ar environment will mitigate the degradation of $CrCl_3$ and $NbSe_2$ thereby, facilitate the fabrication of intricate pristine twisted tunnel junctions.

We also fabricated nominally twisted (0-degree) devices of $NbSe_2$, another prototypical 2D van der Waals superconductor. For $NbSe_2$ flakes we noticed that the yield of re-exfoliation increases if the thickness is > 100 nm. We successfully re-exfoliated $NbSe_2$ flakes of a thickness of 136 nm and 120 nm, whereas the re-exfoliation did not work with the flakes of a lesser thickness (~30 nm). We characterize the fabricated $NbSe_2$ junction in terms of its R vs T and DC I-V response. Figure 6a shows optical micrograph of a $NbSe_2$ junction with gold electrodes deposited through SiN shadow mask. R vs T response (Figure 6b) shows Tc of the junction is 6.2 K, close to the Tc (~6.5 K) of pristine $NbSe_2$[29]. DC I-V characteristic of the junction at 5.7 K is shown in Figure 6c for both directions of bias current sweep. The critical current of the junction at this temperature is 1.35 mA which corresponds to a critical current density of 1.15 $kA/cm^2$; this value of critical current density is comparable to 1.08 $kA/cm^2$ for ~ 6° at 5.7 K to the critical current density report by Farrar et al.[30]

Our experimental results indicate that our method's efficacy in re-exfoliating flakes depends on the vdW material in question. While we successfully demonstrated the re-exfoliation of BSCCO, $CrCl_3$ and $NbSe_2$, our method did not result in the re-exfoliation of other layered materials. Alternatively, it resulted in the complete pick-up and drop of $WSe_2$, $MoS_2$, and $GeBi_2Te_4$ flakes, as illustrated in Figure 5c-e. Amidst these vdW materials, and $GeBi_2Te_4$ are unstable, making our chemical-free assembly particularly advantageous for their fabrication. Numerous recent reports carry out the fabrication of these materials in a glove box, yet despite these efforts, chloroform is still commonly utilized to dissolve the PC/PPC films, resulting in unintended degradation. In contrast, the approach outlined in this manuscript effectively eliminates the need for chemical treatment. Stable TMDCs such as $WSe_2$ and $MoS_2$ also benefit from this cryogenic pick-up method. These materials are often used to assemble heterostructures with highly unstable materials, such as MoTe$_2$, to stabilize their angle, form a moiré lattice, or induce proximity effects[31]. Our low-temperature pick-up/drop method can integrate these stable TMDCs with highly unstable materials to explore exotic moiré physics.

It should also be noted that competition exists between the complete pick-up and re-exfoliation of these layered materials, which is contingent upon a multitude of factors, such as their crystal structure, as well as the competing van der Waals forces at play between the inter-atomic layers, substrate, and PDMS[32,33]. In an effort to re-exfoliate the previously mentioned flakes, which were found to be completely picked up, an attempt was made to enhance the adhesion between the Si/SiO$_2$ substrate and the flake by carrying out a surface treatment process whereby the substrate was exposed to O$_2$ plasma for 10 min at 50 W. Nonetheless, our observations indicate that these materials were still prone to complete pick-up, indicating that re-exfoliation could not be achieved for them. To investigate the impact of flake thickness on the success of our modified cryogenic transfer method, we conducted experiments on flakes with varying thicknesses for all the aforementioned materials. Table 1 summarizes the results obtained for re-exfoliation, pick-up, and drop.

**Table 1:** Summary of cryogenic exfoliation yield for 2D materials with different thicknesses on multiple re-exfoliation attempt

| Materials | Flakes Thickness | Substrate surface treatment | Number of trials | Pick up/cleaving |
|---|---|---|---|---|
| BSCCO | 52 nm | O$_2$ Plasma Cleaned | 1 | Cleaved |
| | 87 nm | | 1 | |
| | 37 nm | | 2 | |
| | 80 nm | | 1 | |
| CrCl$_3$ | ~90 nm | | 1 | Cleaved |
| | ~80 nm | | 2 | |
| | ~30 nm | | 1 | |
| | ~20 nm | | 3 | Not cleaved |
| | ~15 nm | | 3 | |
| NbSe$_2$ | ~ 30 nm | O$_2$ Plasma Cleaned | 5 | Not Cleaved |
| | | UV Ozone | 5 | |
| | ~ 130 nm | O$_2$ Plasma Cleaned | 5 | Not Cleaved |
| | | UV Ozone | 5 | Cleaved |
| MoS$_2$ | ~60 nm | O$_2$ Plasma Cleaned | 3 | Pickup |

|  | ~30 nm |  | 2 |  |
|  | ~5 nm |  | 2 |  |
|  | ~3 nm |  | 1 |  |
| WSe$_2$ | ~60 nm |  | 2 | Pickup |
|  | ~30 nm |  | 1 |  |
|  | ~6 nm |  | 2 |  |
|  | ~3 nm |  | 1 |  |
| GeBi$_2$Te$_4$ | ~50 nm |  | 3 | Pickup |
|  | ~30 nm |  | 2 |  |

**Conclusion**

In summary, we have developed a facile method of re-exfoliation, picking up and transferring 2D materials using a low-temperature cryogenic setup. The developed technique is easy for stacking and twisting 2D materials showing the good quality of the interface and being most suitable for chemically sensitive materials. As an example, we have fabricated 0° twisted device of BSCCO which is cleaved and transferred using this technique. Electrical transport measurements were carried out to verify the high quality of the device. We also fabricated 0° twisted device of NbSe$_2$ and characterized it. In addition, the method is also tried for other systems such as CrCl$_3$ and GeBi$_2$Te$_4$, including TMDCs such as WSe$_2$, MoS$_2$, and NbSe$_2$. In general, the developed method provides a way for micro-fabrication of any chemically sensitive van der Waals materials.

## ACKNOWLEDGMENTS

We acknowledge the Department of Science and Technology (DST), Nanomission grant SR/NM/NS-45/2016, CORE grant CRG/2020/003836, and Department of Atomic Energy (DAE) Government of India for support.

## AUTHOR CONTRIBUTIONS

V.P. and S.G. developed the cryogenic exfoliation setup and fabricated the devices. A.B., K. and A.D. helped in device fabrication. A.S helped in the shadow mask fabrication. V.P., S.G. and A.B. did the measurements. R.K., D.A.J. grew the BSCCO crystals under the supervision of A.T. V.P. and M.M.D. wrote the manuscript with inputs from all authors. M.M.D. supervised the project.

## COMPETING INTERESTS

The authors declare no competing interests.

**References:**

1. Novoselov, K. S.; Mishchenko, A.; Carvalho, A.; Castro Neto, A. H. 2D Materials and van der Waals Heterostructures. *Science* **2016**, 353, aac9439. [DOI: 10.1126/science.aac9439](https://doi.org/10.1126/science.aac9439)

2. Desai, S. B.; Madhvapathy, S. R.; Sachid, A. B.; Llinas, J. P.; Wang, Q.; Ho Ahn, G.; Pitner, G.; Kim, M. J.; Bokor, J.; Hu, C.; Philip Wong, H.-S.; Javey, A. Transistors With 1-nanometer Gate Lengths. *Science* **2016**, 354, 99-102. [DOI: 10.1126/science.aah4698](https://doi.org/10.1126/science.aah4698)

3. Zhu,Y.; Li, Y.; Arefe, G.; Burke, Robert A.; Tan, C.; Hao, Y.; Liu, X.; Liu, X.; Yoo, W. J.; Dubey, M.; Lin, Q.; Hone, J. C. Monolayer Molybdenum Disulfide Transistors With Single-Atom-Thick Gates. *Nano Lett.* **2018,** 18, 3807-3813. [DOI: 10.1021/acs.nanolett.8b01091](https://doi.org/10.1021/acs.nanolett.8b01091)

4. Sinha, S.; Adak, P.C.; Chakraborty, A.; Das, K.; Debnath, K.; Varma Sangani, L. D.; Watanabe, K.; Taniguchi, T.; Waghmare, U. V.; Agarwal, A.; & Deshmukh M M. Berry Curvature Dipole Senses Topological Transition In A Moiré Superlattice. *Nature Physics* **2022,** 18, 765-770. [doi.org/10.1038/s41567-022-01606-y](https://doi.org/10.1038/s41567-022-01606-y)

5. Wang, Z.; Cheon, C-Y.; Tripathi, M.; Marega, G. M.; Zhao, Y.; Ji, H.G.; Macha, M.; Radenovic, A.; Kis A. Superconducting 2D $NbS_2$ Grown Epitaxially by Chemical Vapor Deposition. *ACS Nano.* **2021,** 15, 18403-18410. [doi.org/10.1021/acsnano.1c07956](https://doi.org/10.1021/acsnano.1c07956)

6. Weng, Q.; Li, G.; Feng, X.; Nielsch, K.; Golberg, D.; Schmidt O. G. Electronic and Optical Properties of 2D Materials Constructed from Light Atoms. *Adv. Mater.* **2018,** 30, 1801600. [DOI: 10.1002/adma.2018016001](https://doi.org/10.1002/adma.2018016001)

7. Mootheri, V.; Arutchelvan, G.; Banerjee, S.; Sutar, S.; Leonhardt, A.; Boulon, M. A.; Huyghebaert, C.; Houss, M.; Asselberghs, I.; Radu, I.; Heyns, M.; Lin, D. Graphene Based van der Waals Contacts on $MoS_2$ Field Effect Transistors. *2D Mater.* **2020,** 8, 015003. [DOI 10.1088/2053-1583/abb959](https://doi.org/10.1088/2053-1583/abb959)

8. Huang, J-K.; Pu, J.; Hsu, C-L.; Chiu, M-H.; Juang, Z-Y.; Chang, Y-H.; Chang W-H.; Iwasa, Y.; Takenobu, T.; LL-j. Large-Area Synthesis of Highly Crystalline $WSe_2$ Monolayers and Device Applications. *ACS Nano* **2014,** 8, 923-930. [doi.org/10.1021/nn405719x](https://doi.org/10.1021/nn405719x)

9.  Yang, W.; Chen, G.; Shi, Z.; Liu, C-C.; Zhang, L.; Xie, G.; Cheng, M.; Wang, D.; Yang, R.; Shi, D.; Watanabe, K.; Taniguchi, T.; Yao, Y.; Zhang Y.; Zhang G.; Epitaxial Growth of Single-Domain Graphene on Hexagonal Boron Nitride. *Nat. Mater.* **2013,** 12, 792-797. doi.org/10.1038/nmat3695

10. Huang, Y. *et al.* Universal Mechanical Exfoliation of Large-Area 2D Crystals. *Nat. Commun.* **2020,** 11, 2453. doi.org/10.1038/s41467-020-16266-w

11. Sterpetti, E.; Biscaras, J.; Erb A.; Shukla, A. Comprehensive Phase Diagram of Two-Dimensional Space Charge Doped $Bi_2Sr_2CaCu_2O_{8+x}$ *Nat. Commun.* **2017,** 8, 2060. doi.org/10.1038/s41467-017-02104-z

12. Ghosh, S.; Vaidya, J.; Datta, S.; Pandeya, R. P.; Jangade, D. A.; Kulkarni, R. N.; Maiti, K.; Thamizhavel, A.; Deshmukh M. M. On-Demand Local Modification of High-$T_c$ Superconductivity in Few Unit-Cell Thick $Bi_2Sr_2CaCu_2O_{8+\delta}$ *Advanced Materials* **2020,** 32, 2002220. doi.org/10.1002/adma.202002220

13. Gao, L.; Ni, G-X.; Liu, Y.; Liu, B.; Castro Neto, A. H.; Loh, K. P. Face-To-Face Transfer of Wafer-Scale Graphene Films. *Nature* **2014,** 505, 190-194. doi.org/10.1038/nature12763

14. Castellanos-Gomez, A.; Buscema, M.; Molenaar, R.; Singh, V.; Janssen, L.; van der Zant, H-S-J.; Steele, G. A. Deterministic Transfer of Two-Dimensional Materials by All-Dry Viscoelastic Stamping. *2D Mater.* **2014,** 1, 011002. doi.org/10.1088/2053-1583/1/1/011002

15. Purdie, D. G.; Pugno, N. M.; Taniguchi, T.; Watanabe, K.; Ferrari A. C.; Lombardo, A. Cleaning Interfaces in Layered Materials Heterostructures. *Nature Communications* **2018,** 9, 5387. doi.org/10.1038/s41467-018-07558-3

16. Kinoshita, K.; Moriya, R.; Onodera, M.; Wakafuji, Y.; Masubuchi, S.; Watanabe, K.; Taniguchi, T.; Machida, T. Dry Release Transfer of Graphene and Few-Layer h-BN by Utilizing Thermoplasticity of Polypropylene Carbonate. *npj 2D Mater. Appl.* **2019,** 3, 22. doi.org/10.1038/s41699-019-0104-8

17. Pizzocchero, F.; Gammelgaard, L.; Jessen, B. S.; Caridad, J. M.; Wang, L.; Hone, J.; Bøggild. P.; Booth, T. J. The Hot Pick-Up Technique for Batch Assembly of van der Waals Heterostructures. *Nat. Commun.* **2016,** 7, 11894. doi.org/10.1038/ncomms11894

18. Frank Zhao, S. Y.; Poccia, N.; Panetta, M. G.; Yu, C.; Johnson, J. W.; Yoo, H.; Zhong, R.; Gu, G. D.; Watanabe, K.; Taniguchi, T.; Postolova, S. V.; Vinokur, V. M.; Kim, P. Sign-Reversing Hall Effect in Atomically Thin High-Temperature $Bi_{2.1}Sr_{1.9}CaCu_{2.0}O_{8+\delta}$ Superconductors *Phys. Rev. Lett.* **2019,** 122, 247001. DOI: 10.1103/PhysRevLett.122.247001

19. Novoselov, K. S.; Jiang, D.; Schedin, F.; Geim A. K. Two-Dimensional Atomic Crystals *PNAS* **2005,** 102, 10451-10453 doi.org/10.1073/pnas.0502848102

20. Lee, Y.; Martini, M.; Confalone, T.; Shokri, S.; Saggau, C. N.; Wolf, D.; Gu, G.; Watanabe, K.; Taniguchi, T.; Montemurro, D.; Vinokur, V. M.; Nielsch, K.; Poccia, N. Encapsulating High-Temperature Superconducting Twisted van der Waals Heterostructures Blocks Detrimental Effects of Disorder. *Adv. Mater.* **2023,** 35, 2209135. doi.org/10.1002/adma.202209135

21. Yu, Y.; Ma, L.; Cai, P.; Zhong, R.; Ye, C.; Shen, J.; Gu, G. D.; Chen, X. H.; Zhang, Y. High-Temperature Superconductivity in Monolayer $Bi_2Sr_2CaCu_2O_{8+\delta}$ *Nature* **2019,** 575, 156-163. doi.org/10.1038/s41586-019-1718-x

22. Jiang, D.; Hu, T.; You, L.; Li, Q.; Li, A.; Wang, H.; Mu, G.; Chen, Z.; Zhang, H.; Yu, G.; Zhu, J.; Sun, Q.; Lin, C.; Xiao, H.; Xie, X.; Jiang, M. High-$T_c$ Superconductivity in Ultrathin $Bi_2Sr_2CaCu_2O_{8+x}$ Down to Half-Unit-Cell Thickness by Protection with Graphene *Nat. Commun.* **2014,** 5, 5708. doi:10.1038/ncomms6708

23. Sandilands, L. J.; Shen, J. X.; Chugunov, G. M.; Zhao, S. Y. F.; Ono, S.; Ando, Y.; Burch K. S. Stability of Exfoliated $Bi_2Sr_2D_{yx}Ca_{1-x}Cu_2O_{8+x}$ Studied by Raman Microscopy *Phys. Rev. B* **2010,** 82, 064503. doi.org/10.1103/PhysRevB.82.064503

24. Fratini, M.; Poccia, N.; Ricci, A.; Campi, G.; Burghammer, M.; Aeppli, G.; Bianconi, A. Scale-Free Structural Organization of Oxygen Interstitials in $la_2CuO_{4+y}$ *Nature* **2010,** 466, 841-844. doi.org/10.1038/nature09260

25. Zhao, S. Y. F.; Poccia, N.; Cui, X.; Volkov, P.A.; Yoo, H.; Engelke, R.; Ronen, Y.; Zhong, R.; Gu, G.; Plugge, S.; Tummuru, T.; Franz, M.; Pixley, J. H.; Kim, P. Emergent Interfacial Superconductivity Between Twisted Cuprate Superconductors. **2021,** arXiv:2108.13455. URL http://arxiv.org/abs/2108.13455

26. Wakafuji, Y.; Moriya, R.; Masubuchi, S.; Watanabe, K.; Taniguchi, T.; Machida, T. 3D Manipulation of 2D Materials Using Microdome Polymer *Nano Lett.* **2020,** 20, 2486-2492. doi.org/10.1021/acs.nanolett.9b05228

27. S. Ghosh; V. Patil; A. Basu; Kuldeep; A. Dutta; D. A. Jangade; R. Kulkarni; A. Thamizhavel; J. F. Steiner; F. von Oppen; M. M. Deshmukh; High-temperature Josephson diode *Nature Materials* doi.org/10.1038/s41563-024-01804-4

28. Kleiner, R., Zhou, X., Dorsch, E. et al. Space-time crystalline order of a high-critical-temperature superconductor with intrinsic Josephson junctions. *Nat. Commun.* **2021,** 12, 6038. doi.org/10.1038/s41467-021-26132-y

29. E. Zhang; X. Xu; Yi-Chao Zou; L. Ai; X. Dong; C. Huang; P. Leng; S. Liu; Y. Zhang; Z. Jia; X. Peng; M. Zhao; Y. Yang; Z. Li; H. Guo; S. J. Haigh; N. Nagaosa; J. Shen; F. Xiu; Nonreciprocal superconducting $NbSe_2$ antenna *Nat. Commun.* **2020,** 11, 5634. doi.org/10.1038/s41467-020-19459-5

30. L. S. Farrar; A. Nevill; Z. J. Lim; G. Balakrishnan; S. Dale; S. J. Bending; Superconducting Quantum Interference in Twisted van der Waals Heterostructures *Nano Lett.* **2021**, 21, 6725. doi.org/10.1021/acs.nanolett.1c00152

31. Zhao, W.; Shen, B.; Tao, Z.; Han, Z.; Kang, K.; Watanabe, K.; Taniguchi, T.; Mak, K. F.; Shan, J. Gate-Tunable Heavy Fermions in A Moiré Kondo Lattice *Nature* **2023,** 616, 61-65. doi.org/10.1038/s41586-023-05800-7

32. Kim, J-Y.; Ju, X.; Ang, K. W.; Chi, D. Van der Waals Layer Transfer of 2D Materials for Monolithic 3D Electronic System Integration: Review and Outlook *ACS Nano* **2023,** 17, 1831-1844. doi.org/10.1021/acsnano.2c10737

33. Zhang, H-Z.; Wu, W-J.; Zhou, L.; Wu, Z.; Zhu, J.; Steering on Degrees of Freedom of 2D van der Waals Heterostructures *Small Sci.* **2022,** 2, 2100033. DOI: 10.1002/smsc.202100033

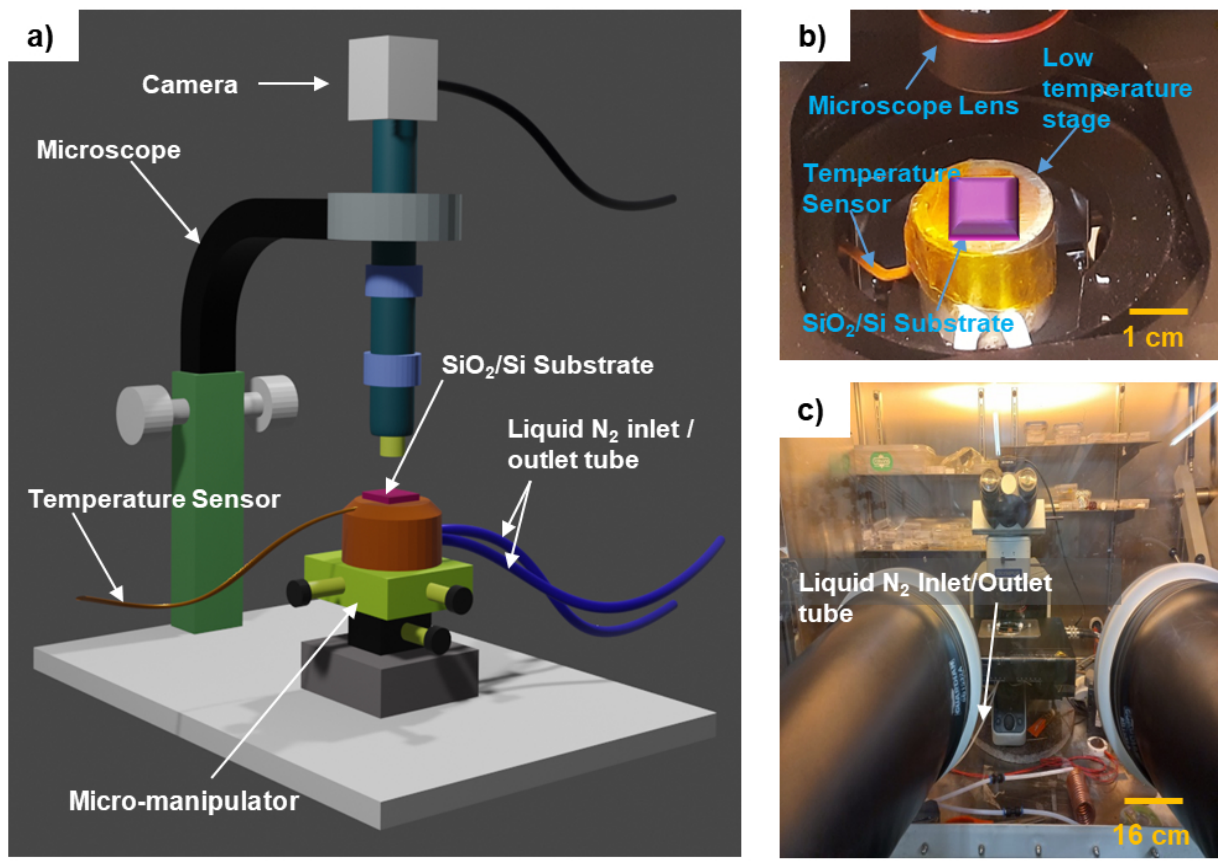

**Figure 1:** (a) Schematic of the cryogenic exfoliation setup (b) Photograph of the low temperature stage (c) Photograph of the cryogenic exfoliation setup inside the glove box.

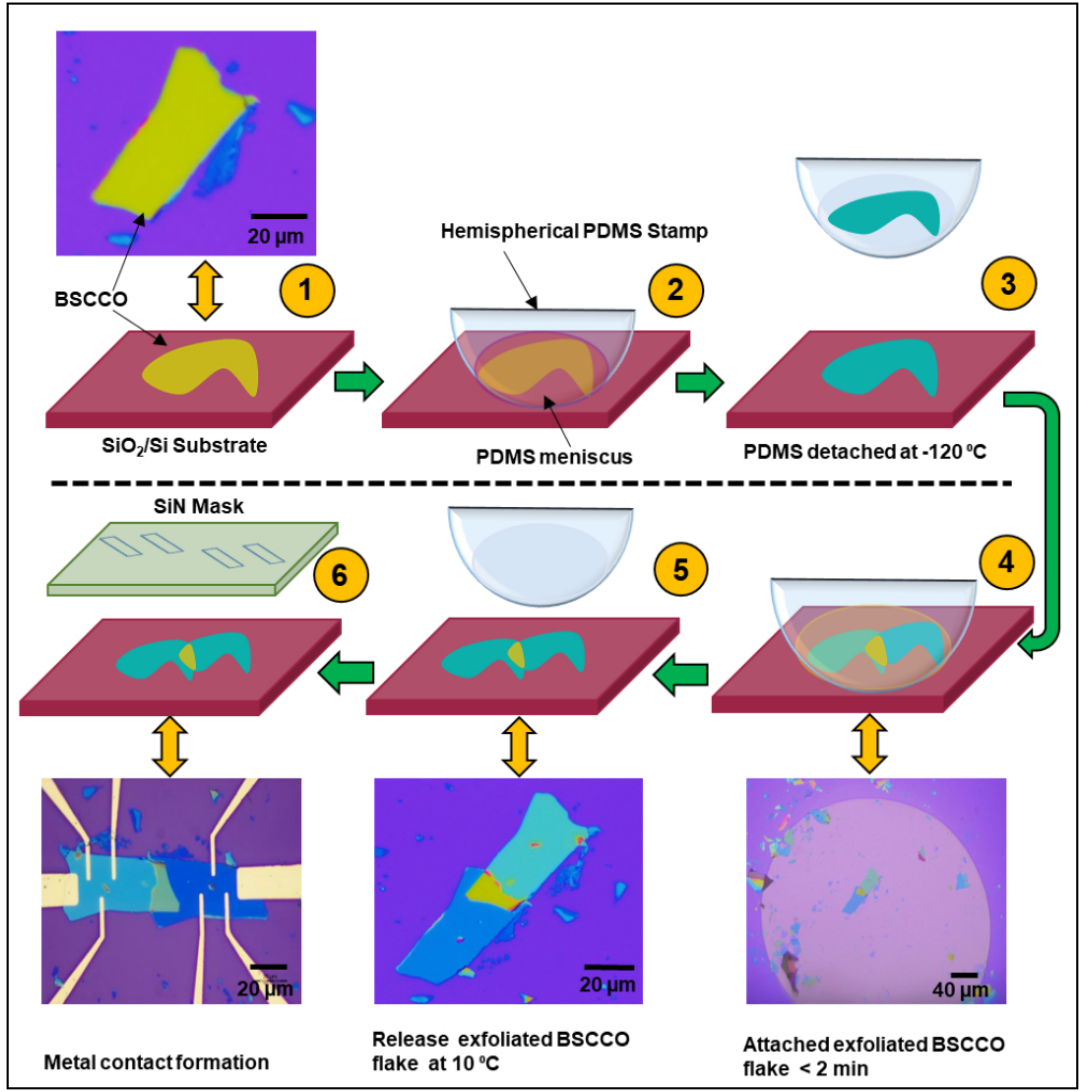

**Figure 2:** Low-temperature fabrication process for twisted BSCCO devices. Step 1: Exfoliate the BSCCO flake on the SiO$_2$/Si substrate and identify it under a microscope. Step 2: Attach the hemispherical PDMS stamp at -25 °C to the exfoliated flake. Step 3: Cool down the stage to -120 °C where the PDMS stamp enters into the glass transition stage, causing the re-exfoliation of BSCCO flake. Step 4: Quickly rotate the PDMS stamp to the desired angle and attach it to the flake of the substrate. Step 5: Warm up the stage to 10 °C and remove the PDMS stamp to realize a BSCCO junction on the SiO$_2$/Si substrate. Step 6: Align the SiN mask with the fabricated stack and deposit the electrodes through it via evaporation. In steps 1, 4, 5, and 6, bidirectional arrows show the schematic and optical images of the corresponding steps.

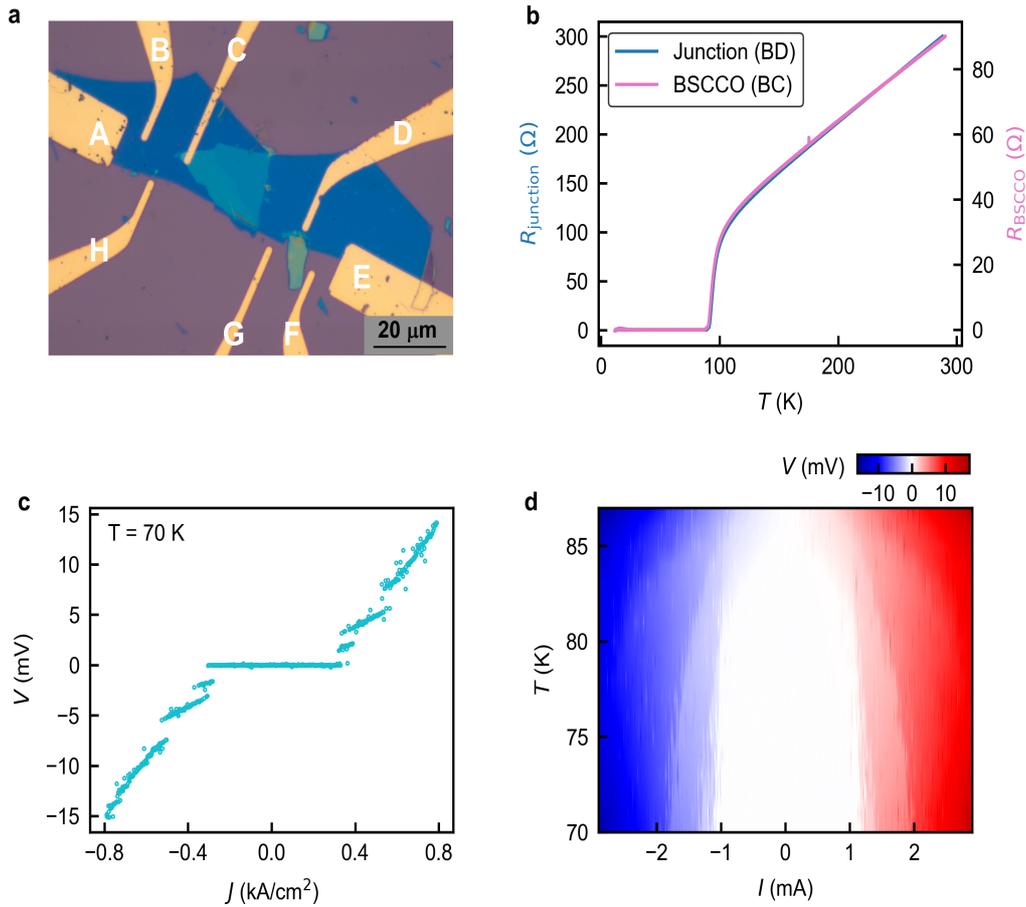

**Figure 3:** (a) Optical image of a 0° twisted BSCCO device. (b) Corresponding resistance (R) versus temperature (T) data for the junction (BD electrode) and pristine BSCCO flake (BC electrode) of 0° twisted device. (c) Analogous current density versus voltage (*J-V*) characteristics of the 0° twisted device measured at 70 K. (d) A colour scale plot of dc *I-V* characteristics of the 0° twisted device as a function of temperature. The white-shaded dome area is the superconducting transport regime of the device.

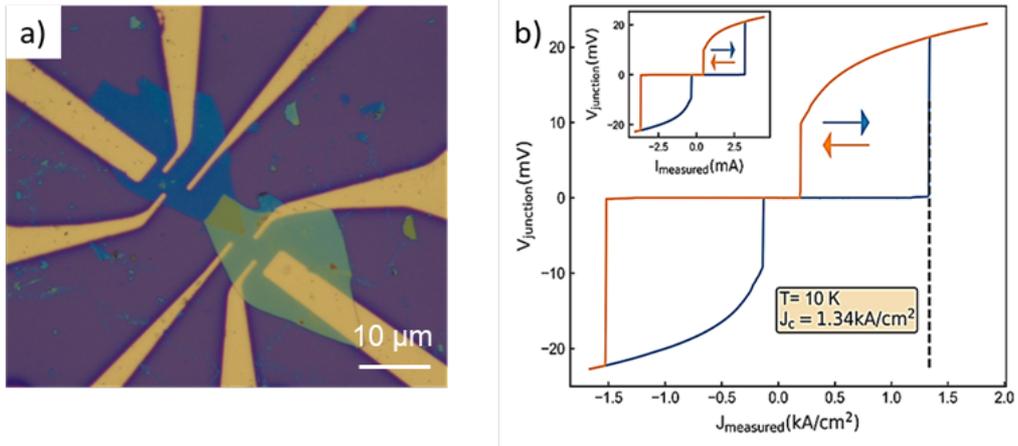

**Figure 4:** (a) Optical image of a 0° twisted BSCCO device. (b) Current density versus voltage (J-V) characteristics of the 0° twisted BSCCO device measured at 10 K, in the inset the I-V characteristic of the device is shown for the same temperature.

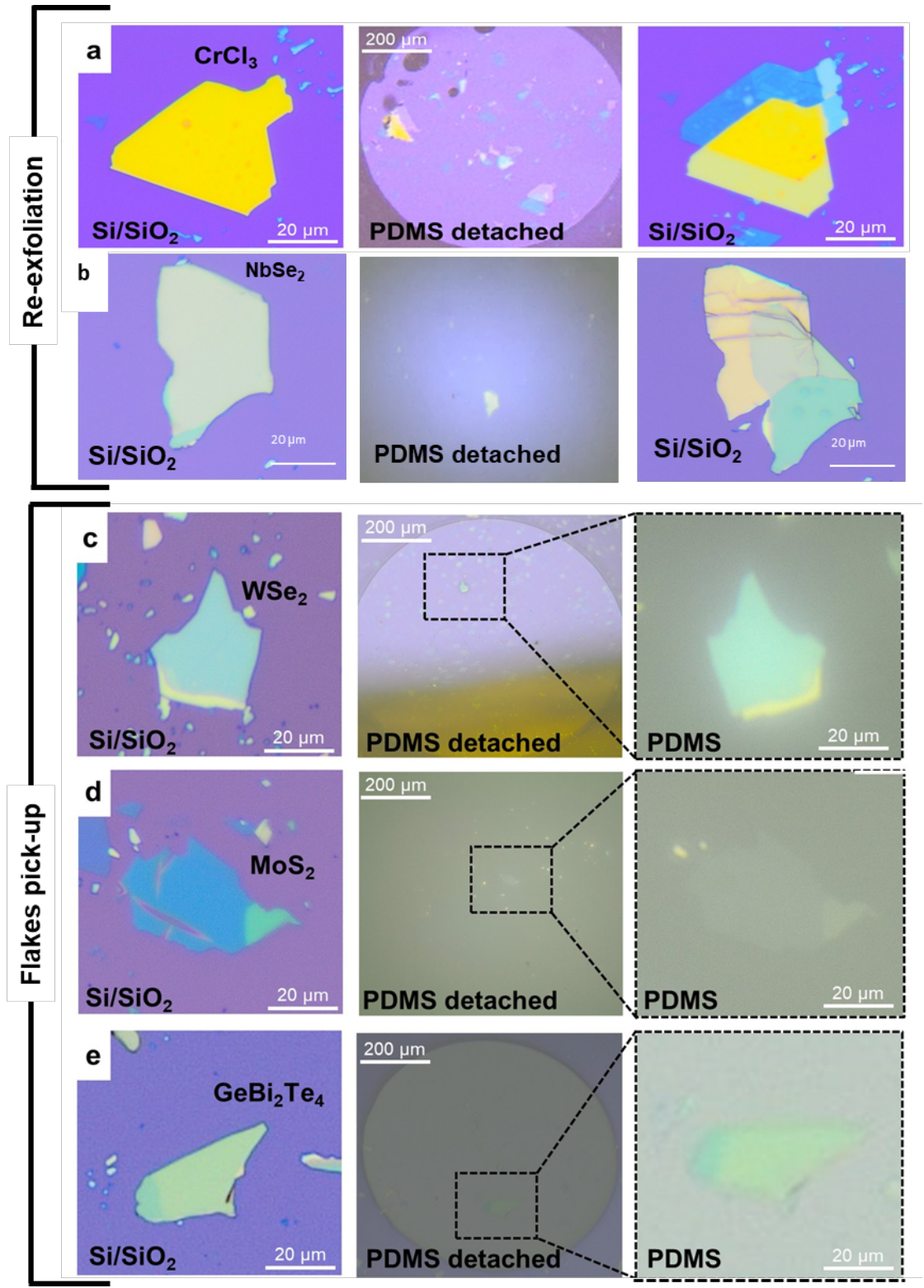

**Figure 5:** (a and b) Re-exfoliation of $CrCl_3$ and $NbSe_2$: Optical image of the $CrCl_3$ and $NbSe_2$ flakes, followed by its re-exfoliation and pick up on the detached PDMS stamp. (b-e) Pick-up of other vdW materials: Optical images of the thick flakes before pick-up on $SiO_2$/Si substrate, followed by their pick-up on detached PDMS stamp: (b) $NbSe_2$ (c) $WSe_2$ (d) $MoS_2$ (e) $GeBi_2Te_4$.

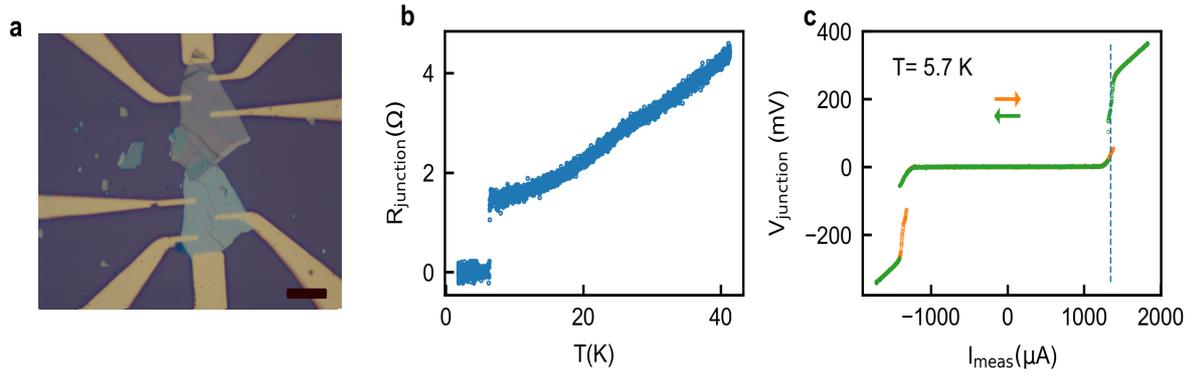

**Figure 6:** (a) Optical image of twisted NbSe$_2$ device. (b) RT characterization of 0° twisted NbSe2 device (c) DC IV characteristic of the device measured at 5.7 K.